\documentclass[sigconf]{acmart}
\AtBeginDocument{%
  \providecommand\BibTeX{{%
    \normalfont B\kern-0.5em{\scshape i\kern-0.25em b}\kern-0.8em\TeX}}}

\copyrightyear{2022}
\acmYear{2022}
\setcopyright{rightsretained}
 \acmConference[AIES'22]{Proceedings of the 2022 AAAI/ACM Conference on AI, Ethics, and Society}{August 1--3, 2022}{Oxford, United Kingdom}
\acmBooktitle{Proceedings of the 2022 AAAI/ACM Conference on AI, Ethics, and Society (AIES'22), August 1--3, 2022, Oxford, United Kingdom}\acmDOI{10.1145/3514094.3534136}
\acmISBN{978-1-4503-9247-1/22/08}

\usepackage{etoolbox}
\makeatletter
\patchcmd{\maketitle}{\@copyrightpermission}{
   \begin{minipage}{0.3\columnwidth}
     \href{https://creativecommons.org/licenses/by-nc-sa/4.0/}{\includegraphics[width=0.90\textwidth]{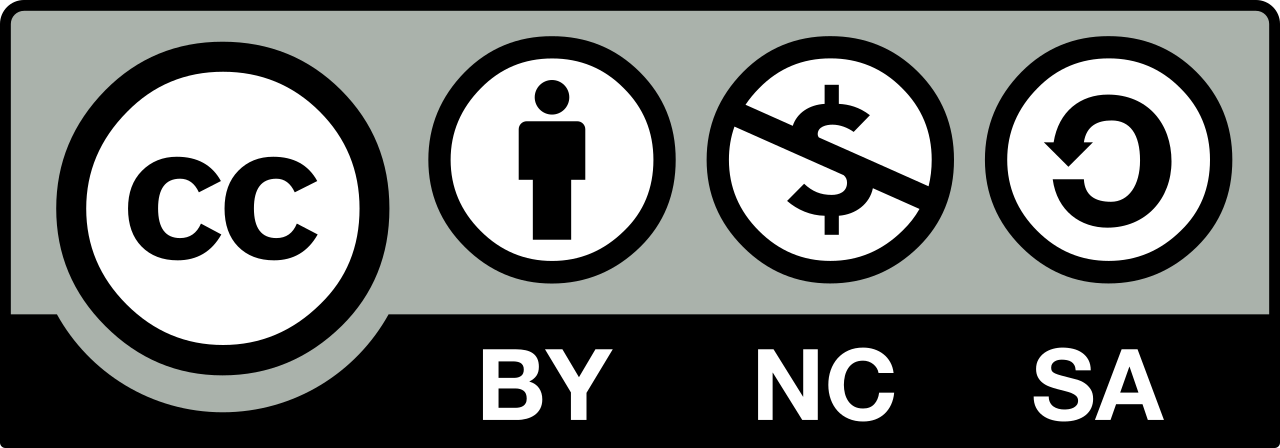}}
   \end{minipage}\hfill
   \begin{minipage}{0.7\columnwidth}
     \href{https://creativecommons.org/licenses/by-nc-sa/4.0/}{This work is licensed under a Creative Commons Attribution-NonCommercial-ShareAlike International 4.0 License.}
   \end{minipage}

   \vspace{5pt}
}{}{}

\makeatother

\usepackage{tikz}
\usepackage{pgfplots}
\usepackage{pgfplotstable}
\usetikzlibrary{patterns}
\usepgfplotslibrary{colormaps}
\usetikzlibrary{pgfplots.colormaps}
\usepackage{pgf-pie} 
\usepackage{xcolor,colortbl}
\usepackage{multirow}
\usepackage{xurl}

\usepackage{makecell}

\pgfplotsset{
        colormap={test}{[2pt]
            rgb255(0pt)=(48, 133, 252);
            rgb255(500pt)=(175, 4, 201);
            rgb255(1000pt)=(242, 41, 10);
        },
    }

\begin{document}
\fancyhead{}

\title{American == White in Multimodal Language-and-Image AI}

\author{Robert Wolfe}
\affiliation{%
  \institution{University of Washington}
    \institution{Information School}
  \city{Seattle}
  \state{WA}
  \country{USA}
  }
\email{rwolfe3@uw.edu}

\author{Aylin Caliskan}
\affiliation{%
  \institution{University of Washington}
    \institution{Information School}
  \city{Seattle}
  \state{WA}
  \country{USA}
  }
\email{aylin@uw.edu}


\renewcommand{\shortauthors}{Anonymous}

\begin{abstract}
  Three state-of-the-art language-and-image AI models, CLIP, SLIP, and BLIP, are evaluated for evidence of a bias previously observed in social and experimental psychology: equating American identity with being White. Embedding association tests (EATs) using standardized images of self-identified Asian, Black, Latina/o, and White individuals from the Chicago Face Database (CFD) reveal that White individuals are more associated with collective in-group words than are Asian, Black, or Latina/o individuals, with effect sizes $>.4$ for White vs. Asian comparisons across all models. In assessments of three core aspects of American identity reported by social psychologists, single-category EATs reveal that images of White individuals are more associated with patriotism and with being born in America, but that, consistent with prior findings in psychology, White individuals are associated with being less likely to treat people of all races and backgrounds equally. Additional tests reveal that the number of images of Black individuals returned by an image ranking task is more strongly correlated  with state-level implicit bias scores for White individuals (Pearson's $\rho=.63$ in CLIP, $\rho = .69$ in BLIP) than are state demographics ($\rho = .60$), suggesting a relationship between regional prototypicality and implicit bias. Three downstream machine learning tasks demonstrate biases associating American with White. In a visual question answering task using BLIP, 97\% of White individuals are identified as American, compared to only 3\% of Asian individuals. When asked in what state the individual depicted lives in, the model responds China 53\% of the time for Asian individuals, but always with an American state for White individuals. In an image captioning task, BLIP remarks upon the race of Asian individuals as much as 36\% of the time, but never remarks upon race for White individuals. Finally, provided with an initialization image from the CFD and the text "an American person," a synthetic image generator (VQGAN) using the text-based guidance of CLIP lightens the skin tone of individuals of all races (by 35\% for Black individuals, based on pixel brightness). The results indicate that  biases equating American identity with being White are learned by language-and-image AI, and propagate to downstream applications of such models.
\end{abstract}



\maketitle

\section{Introduction}

The United States is a multiethnic and pluralistic country whose citizens espouse a commitment to the principles of equality and inclusion in the American identity for all people, regardless of race or ethnicity \cite{schuman1997racial,gross1999citizenship}. Yet, as in many societies which outwardly express a commitment to the ideals of equality, the structure of American society distributes opportunities unequally \cite{o2020structural}, and experimental psychologists have found that, even among Americans who affirm a belief in equality for all races, to be American is implicitly associated with being White \cite{devos2005american}.

Artificial Intelligence (AI) is increasingly used to determine access to the benefits, responsibilities, and opportunities which attend life as an American. Computer vision, in particular, is employed to monitor and police American territorial borders and ports of entry \cite{phippen2021smartborder}, and but for bipartisan public outcry \cite{rappeport2022facialrecognition} informed by the research of \citet{buolamwini2018gender}, the U.S. Internal Revenue Service (IRS) would have required American citizens to use third-party facial recognition AI to create an online IRS account \cite{irs2022transition}. While most deployed systems currently use supervised computer vision models designed to recognize a predefined set of image classes, the field of computer vision underwent a transformation in early 2021 with the introduction of CLIP (“Contrastive Language Image Pretraining”) \cite{radford2021learning}. CLIP is the first practical "zero-shot" language-and-image model, a system which learns to match images to descriptive text, and which performs competitively with state-of-the-art supervised models without ever explicitly training on computer vision evaluation datasets \cite{radford2021learning}. CLIP and its successors allow for the definition of image classes in natural language, and have been adapted for numerous domains, including analysis of satellite imagery \cite{radford2021learning}, zero-shot object detection \cite{gu2021open}, and text-based guidance of synthetic image generators \cite{ramesh2021zero,nichol2021glide}. Yet natural language supervision renders computer vision susceptible not only to visual biases, but to biases in human language: \citet{goh2021multimodal} find that neurons in the CLIP image encoder associate Muslims with terrorism, while \citet{wang2021gender} show that images of men are over-represented in occupational search queries when using CLIP for image retrieval.

\begin{figure*}
    \centering
    \includegraphics[width=\textwidth]{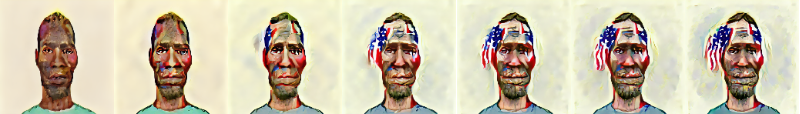}
    \caption{\small Provided with an initialization image of a person who self-identifies as Black, a CLIP-guided image generator (VQGAN) prompted to create "an American person" lightens the skin tone of the output image. Images are from iterations 0 through 175 in steps of 25.}
    \label{fig:gan_change_b_w}
    \Description{Figure depicting lightening of skin tone by VQGAN-CLIP to match the input text "an American person."}
\end{figure*}

The advances realized by CLIP offer an opportunity to study the propagation to AI of a bias previously observed in experimental psychology: specifically, bias associating American identity with being White \cite{devos2005american}. Motivated by the prior discovery of implicit biases in artificial intelligence \cite{caliskan2017semantics} and the consequential nature of this bias for mediating the experiences and opportunities afforded to people for whom American identity is conferred or withheld \cite{armenta2013you,huynh2015boundaries}, the present research examines three English-language language-and-image AI models, CLIP \cite{radford2021learning}, SLIP ("Self-Supervised Language-Image Pretraining") \cite{mu2021slip}, and BLIP ("Bootstrapping Language-Image Pretraining") \cite{li2022blip}, for biases associating American identity with being White, drawing on the work of \citet{devos2005american} to inform the methodology. The primary contributions of the research follow below. Research code is public at \url{https://github.com/wolferobert3/american_white}.

\begin{enumerate}
    \item \textbf{Visual semantic\footnote{When the term "visual semantic" is used in this research, it refers specifically to a joint vision-and-language embedding space. The broader class of models which form language and image representations and may use them in downstream tasks are referred to as language-and-image AI.} embedding spaces exhibit correspondence with the demographics of U.S. states, with $R^{2}$ coefficient of up to .74, in CLIP.} An image ranking task uses the photos of the Chicago Face Database (CFD), a database of standardized images of self-identified Asian, Black, Latina/o, and White individuals intended for the psychological study of race \cite{ma2015chicago}. The images most associated with the text prompt “a photo of someone who lives in the state of [state]” are returned for each of the fifty U.S. states and the District of Columbia. Results correspond to statistics collected in the 2019 U.S. census, with R2 coefficient of .74 for CLIP, .54 for BLIP, and .33 for SLIP, corresponding to the size of each model's training data (400 million, 129 million, and 15 million image-text pairs, respectively). However, language-and-image AI tends to overestimate the population of Black, Asian, or Latina/o individuals living in a state, indicating that location-based categorical associations in AI are not the straightforward result of “learning” the population of a state, but reflect biases expressed in human descriptions. Results also reveal that the proportion of images of Black individuals returned correlates positively with the implicit racial bias of White individuals living in a state, and that the correlation is stronger than correlations between human implicit bias and population statistics. This suggests that implicit bias is related to the threat to the prototypicality of White individuals, as measured via representation in AI.
    \item \textbf{Visual semantic embedding spaces associate American with White.} Two language-image association tests are adapted from the work of \citet{devos2005american}. The first is a We vs. They embedding association test (EAT), testing association with collective in-group words vs. out-group words. CFD images of people who self-identify as Asian, Black, Latina/o, and White are the target groups of the EAT, for which the differential association is tested. Across three models, images of White individuals are consistently associated with collective in-group words when compared with images of Asian, Black, or Latina/o individuals, with all effect sizes for the White vs. Asian comparison $> 0.4$. A second EAT measures association with three characteristics found by \citet{devos2005american} to best describe what American identity according to study participants: patriotism, nativism (being born in the U.S.), and egalitarianism (in this case, treating people of all races and backgrounds equally). Across models, White individuals are differentially associated with patriotism and with nativism. However, White individuals are strongly disassociated from egalitarianism. These findings in AI reflect those of \citet{devos2005american}.
    \item \textbf{Language-and-Image AI associates American identity with being White in three downstream machine learning tasks: visual question answering, image captioning, and text-guided synthetic image generation.} Prompted with the question “Is this person an American?”, the BLIP visual question answering head answers "Yes" 97\% of the time for White individuals, 68\% of the time for Black individuals, 61\% of the time for Latina/o individuals, and only 3\% of the time for Asian individuals. When prompted with the question “What state does this person live in?”, the BLIP visual question answering head answers “China” 53\% of the time for Asian individuals, while responding with the name of a U.S. state more than 97\% of the time for all other images. Prompted to automatically generate a caption for each photo in the Chicago Face Database, the BLIP image captioning head never remarks on the race or ethnicity of White individuals, but remarks on the race or ethnicity of Asian individuals up to 36\% of the time, and up to 18\% of the time for Black individuals. The race or ethnicity of White individuals is never described. Prompted to generate an image of “an American person,” a synthetic image generator guided by the features of CLIP produces images of white individuals with blonde hair. As seen in Figure \ref{fig:gan_change_b_w}, when provided with an initialization image of an individual from the CFD, the CLIP-guided image generator changes the color of the individual’s skin to white. For initialization images of Black individuals, mean pixel brightness for a 50x50 pixel crop of the forehead increases by 35\% after 80 training iterations, reflecting lightening of skin tone.
\end{enumerate}

The present research indicates that language-and-image AI associates American with White in ways that reflect human biases through similar mechanisms of association. This research also indicates that biases of association with American identity observed in the embedding spaces of language-and-image models influence biases in downstream applications of vision and language AI.

\section{Related Work}

Relevant work concerning associations of American with being White, racial bias in AI, and language-and-image AI is reviewed.

\noindent\textbf{American Identity Bias} This research is grounded in foundational work in experimental psychology by \citet{devos2005american}, who observe that, despite explicit affirmation from study participants that members of different ethnic groups should be treated equally, the category "American" is implicitly synonymous with being White for American subjects. The six studies demonstrating this bias included Implicit Association Tests (IATs, which measure unconscious or "implicit" biases and associations in human subjects \cite{greenwald1998measuring}) showing that participants more easily paired White faces with American symbols and landmarks than they did Asian faces, and IATs showing that participants more easily paired the names of White Europeans (not Americans) with American symbols than they did the names of African Americans and Asian Americans \cite{devos2005american}. Subsequent research affirms that a societal in-group may make itself synonymous with a larger human category. \citet{wenzel2008superordinate} find that members of an ethnocentric in-group tend to generalize their characteristics onto a positively valued superordinate category to increase the prototypicality of the in-group, and that out-group differences from the prototypical in-group norm are evaluated as negative deviations. \citet{danbold2015no} find that fear of losing such prototypical status in America predicts White Americans' greater support for assimilation, and lower support for diversity. 

Other psychological research has sought to model the interaction of racial status and perceived foreignness in the U.S. \citet{zou2017two} survey Asian, Black, Latina/o, and White American subjects regarding their recent and overall experiences with racial prejudice in the U.S., and model U.S. racial dynamics using a quadrant of perceived superiority vs. inferiority and perceived American identity vs. foreignness, wherein White Americans are perceived and treated as superior and American; Black Americans are perceived and treated as inferior and relatively American; Latina/o Americans are perceived and treated as inferior and foreign; and Asian Americans are perceived and treated as foreign but relatively superior to Black and Latina/o Americans. Recent work finds that ethnocentric biases may change along with regional demographics: \citet{devos2021temporal} finds that steeper linear increases in the proportion of Asian Americans living in a metropolitan area over time are associated with greater implicit inclusion in national identity.

Research in psychology also finds that ethnocentric American biases have consequences for the mental health and opportunities afforded to Asian Americans and Latina/o Americans. \citet{armenta2013you} find that, among U.S.-born Asian Americans and Latina/o Americans, perceived objectification as a foreigner correlates with lower life satisfaction, greater depressive symptoms, and lower self-esteem. \citet{huynh2011perpetual} find that awareness of the perpetual foreigner stereotype, which asserts that people who are not White will always be seen as foreign in the U.S., predicts lower sense of belonging to American culture, lower hope and life satisfaction for Asian Americans, and greater incidence of depression for Latina/o Americans. \citet{yogeeswaran2010will} find that the stronger an individual's implicit bias associating White with the prototypical American, the less willing the individual is to hire Asian Americans in national security jobs. \citet{huynh2015boundaries} find a relationship between White prototypicality and antiminority policy attitudes and acculturation ideologies in a study of White Americans.

The present research adapts two of the studies of \citet{devos2005american} to examine biases in language-and-image AI. The first is a We/They IAT, wherein one group of words (we, our, ourselves) implies a collective in-group identity, while another group of words (they, other, themselves) implies out-group identity \cite{devos2005american}. \citet{devos2005american} find that it is easier for participants to pair in-group words with the faces of individuals with whom they share an ethnic identity. The second study of \citet{devos2005american} adapted in this work is drawn from a survey of participants to assess the explicit attributes most associated with American identity. \citet{devos2005american} find that three attributes are most associated with American identity: egalitarianism, or treating people of all races and backgrounds equally; patriotism; and native status, or being born in America. Of these attributes, Asian Americans were perceived as the least likely to be patriotic, and the least likely to have been born in America, while White Americans were perceived as being less egalitarian than Asian Americans or African Americans \cite{devos2005american}.

\noindent\textbf{Bias in AI} Prior research on racial bias in AI has addressed, primarily, four prongs: failures of generalization of state-of-the-art technology \cite{buolamwini2018gender}; under-representation in machine learning datasets \cite{dodge2021documenting,wolfe2021low}; biases of association reflected in embedding spaces \cite{caliskan2017semantics,guo2021detecting}; and downstream biases in applications of AI \cite{pandey2021disparate}. The current research examines questions of bias and identity both in embedding spaces and in downstream tasks demonstrating disparate impact. 

\noindent\textbf{\textit{Implicit Bias and the Word Embedding Association Test}} A significant component of the present research is the adaptation of methods grounded in experimental psychology to observe biases in machine learned representations. Foundational research on humanlike biases in word embeddings was contributed by \citet{caliskan2017semantics}, who introduced the Word Embedding Association Test (WEAT), an adaptation of the IAT of \citet{greenwald1998measuring} which showed that machine-learned semantics derived from internet-scale web corpora reflect the biases of the populations who produce them. The WEAT measures the differential angular similarity between two groups of target words with two groups of attribute words, and returns an effect size (Cohen's $d$ \cite{cohen1992statistical}) and a $p$-value indicating statistical significance. The single-category SC-WEAT measures the differential angular similarity of a single word with two attribute groups. The WEAT and SC-WEAT allow for the measurement of association with concepts, and subsequent work extends the WEAT to contextualized word embeddings \cite{guo2021detecting,wolfe2022vast} and sentence embeddings \cite{may2019measuring} in language models such as BERT \cite{devlin-etal-2019-bert} and GPT-2 \cite{radford2019language}. The appendix includes formulae for the WEAT and SC-WEAT.

\noindent\textbf{\textit{Impact of Training Data}} The composition of machine learning training datasets has been shown to have significant impacts on the biases learned by a model. \citet{brunet2019understanding} show that the least frequently occurring words in static word embedding training corpora are also the most biased, and have the most unstable representations. \citet{wolfe2021low} find that names belonging predominantly to Asian, Black, and Latina/o Americans are underrepresented in the training corpora of language models, leading to bias and overfitting in the pretrained model.  \citet{dodge2021documenting} find that methods intended to prevent bias in constructing the C4 corpus also remove data created by and about marginalized populations. \citet{caliskan2022gender} show that word embeddings trained on internet-scale corpora reflect a masculine default which pervades the embedding space.

\noindent\textbf{\textit{Bias in Computer Vision}} Both semantic biases and biases related to failures of generalization for underrepresented populations have been observed previously in computer vision. \citet{buolamwini2018gender} find that the underrepresentation in computer vision datasets results in facial recognition models which disproportionately fail to recognize the faces of women with darker skin. \citet{wilson2019predictive} finds that state-of-the-art object detection systems also fail for people with darker skin. \citet{rhue2018racial} observes that emotion detection systems are more likely to ascribe negative emotions to Black individuals, while \citet{kim2021age} find that emotion detection systems fail to generalize for images of older adults. In accordance with this finding, \citet{park2021understanding} show that computer vision datasets systematically underrepresent older adults. \citet{steed2021image} find that the embedding structure of generative image models such as Image GPT \cite{chen2020generative} is reflective of humanlike social biases, and that the model generates stereotypically sexualized images of women.

\noindent\textbf{\textit{Reflection of Human Society to AI}} In introducing the SC-WEAT, \citet{caliskan2017semantics} demonstrate a linear relationship between the SC-WEAT association of the name of a profession with female attribute words and the proportion of women employed in the profession, with Pearson's $\rho=.88$. The veridical properties of static word embeddings have also rendered them a useful tool for studying human societies. \citet{kozlowski2019geometry} use the geometry of static word embeddings trained on Google N-grams over decades of the twentieth century to show that the material markers used to signify social class changed with the economic transformations of the century. \citet{walter2021diachronic} use diachronic word embeddings trained on German parliamentary proceedings to study the evolution of German political biases over time. \citet{joseph2020word} show that word embeddings can capture population-level beliefs which correspond to the results of surveys of that population. That survey results correlate with bias in embedding spaces is notable for the current research, which adapts results from a survey designed by \citet{devos2005american} to test for biases in an embedding space.

\noindent\textbf{Language-and-Image AI} This research examines bias in three language-and-image AI models: CLIP, SLIP, and BLIP.

\noindent\textbf{\textit{Language-Image Pretraining}} CLIP was the first in a new generation of multimodal language-and-image models trained using natural language supervision \cite{radford2021learning}. Where supervised computer vision trains on a defined set of image classes and associated images, CLIP instead learns to pair text captions collected from the internet with their associated images \cite{radford2021learning}. While the objective is straightforward, the results introduced a new paradigm in computer vision, wherein a user of the pretrained model can define their own image classes in natural language, and retrieve, rank, or classify images based on association with the text \cite{radford2021learning}. In this sense, CLIP is the first "zero-shot" image associator, and is not dependent on the classes and images of any specific dataset \cite{radford2021learning}. The architecture of CLIP is composed of a contextualizing language model (a smaller version of the causal language model GPT-2 \cite{radford2019language}, based on the transformer architecture of \citet{vaswani2017attention}) used to form sentence embeddings, and an image encoder, either a Vision Transformer \cite{dosovitskiy2020image} or a ResNet \cite{he2016deep}. The language and vision models are jointly pretrained, and the representations formed by each are projected into a joint "visual semantic" embedding space \cite{radford2021learning}, wherein cosine similarity is used to assess the similarity between embedded image and embedded text. This research examines the CLIP-ViT-Base-Patch32 model available via the Transformers library of \citet{wolf-etal-2020-transformers}, the most downloaded model of the versions available via Transformers.

SLIP adopts the architecture and training design of CLIP and adds data augmentation to the CLIP objective \cite{mu2021slip}. SLIP randomly resizes and crops images such that they are between 50\% and 100\% the size of the original image, a technique intended to improve the robustness of the model for extracting the semantic content of an image \cite{mu2021slip}. SLIP also adds an image-based self-supervision branch, wherein the model is trained to represent different views of the same image with similar vectors \cite{mu2021slip}. This research examines the ViT-Base version of SLIP trained for 100 epochs, the best performing version of the Base model on zero-shot ImageNet evaluation \cite{mu2021slip}.

BLIP is a multimodal language-and-image encoder-decoder model \cite{li2022blip}. Like CLIP and SLIP, BLIP trains a visual semantic embedding space using contrastive loss to align text and image representations \cite{li2022blip}. Unlike CLIP and SLIP, BLIP is also trained for language-and-image tasks which require text generation, including visual question answering and image captioning, on which it set new state of the art \cite{li2022blip}. BLIP introduced a new synthetic caption generation and filtering technique, known as CapFilt, which filters out noisy or uninformative captions during training \cite{li2022blip}. This research uses the ViT-Base version of BLIP trained on 129 million image-text pairs with MS-COCO fine-tuning for evaluating embedding space associations; the ViT-Base checkpoint with CapFilt-L for automatic image captioning; and the ViT-Base checkpoint with CapFilt-L for visual question answering \cite{li2022blip}. These are the default checkpoints used in a publicly available version of the model \cite{li2022blip}.

\noindent\textbf{\textit{Visual Semantic AI for Guiding and Training Generative Models}} Among the first uses of CLIP was to train the text-to-image generation model DALL-E \cite{ramesh2021zero}, and CLIP subsequently used in the training of GLIDE \cite{nichol2021glide} and DALL-E 2 \cite{ramesh2022hierarchical}, diffusion-based text-to-image generation models. Such models use CLIP representations as ground truth, and are likely to inherit its biases. Because no version of DALL-E or GLIDE capable of generating human images is publicly available, this research examines a CLIP-guided VQGAN \cite{esser2021taming}, which uses convolutional neural networks to learn a vocabulary of image components, and transformers to compose them in high-resolution images \cite{esser2021taming}. VQGAN-CLIP uses the cosine similarity between CLIP-embedded text and VQGAN-generated images as a loss to increase the image's similarity with the target text \cite{crowson2022vqgan}.

\noindent\textbf{\textit{Bias Specific to Language-and-Image AI}} \citet{radford2021learning} and \citet{agarwal2021evaluating} find that CLIP prefers text highlighting the physical features of women, and is more likely to misclassify Black individuals into animal categories. \citet{wolfe2022evidence} provide evidence that CLIP associates images of multiracial individuals with race or ethnicity labels according to a rule of hypodescent, or one-drop rule. \citet{wang2021gender} mitigate gender biases in CLIP image retrieval results by muting gendered neurons in CLIP image embeddings, and by preferentially sampling images of women.  \citet{wolfe2022markedness} find that CLIP draws attention to the race, gender, and age of underrepresented individuals, while leaving these characteristics unmarked for white, male, and middle-aged individuals.

\section{Data}

This research uses the Chicago Face Database (CFD) to evaluate the association of American identity with being White. The training datasets for CLIP, SLIP, and BLIP are also discussed.

\noindent\textbf{The Chicago Face Database}
The CFD is a dataset of images used to study race and ethnicity in psychology \cite{ma2015chicago}. The CFD includes $597$ high-resolution ($2,444$ x $1,718$ pixel) images of male and female volunteers who provided information regarding their self-identified race or ethnicity \cite{ma2015chicago}. The self-identified races and ethnicities reported for the CFD include Asian, Black, Latina/o, and White \cite{ma2015chicago}. CFD images position subjects facing the camera against a white background, such that every subject's face occupies the same area of the image. All subjects are captured with a "neutral" facial expression, and a subset with "happy (open mouth)," "happy (closed mouth)," "angry," and "fearful" facial expressions. In accordance with the methodology of \citet{devos2005american}, this research uses only images of subjects with a neutral facial expression. Because the present research examines biases related to American identity, it is noteworthy that all CFD subjects were recruited in the U.S.

\noindent\textbf{U.S. Census Data} Results for an image ranking experiment are compared to 2019 U.S. state-level census population estimates available at \url{https://www.census.gov/data/tables/time-series/demo/popest/2010s-state-detail.html}. 

\noindent\textbf{IAT Data} This research examines the correlation of the racial association of U.S. states in language-and-image AI with the mean IAT effect sizes for those states. The IAT data for this analysis is obtained from the Project Implicit and is based on eight years of state-level data obtained via the online IAT \cite{ratliff2020documenting}.

\noindent\textbf{Training Data for Language-and-Image AI} The training data for AI models determines to a large extent the biases learned \cite{caliskan2017semantics,wolfe2021low}. The training datasets for CLIP, SLIP, and BLIP are discussed below.

\noindent\textbf{\textit{CLIP}} \citet{radford2021learning} train CLIP on the WebImageText corpus (WIT), a web scrape composed of $400$ million images and associated captions. \citet{radford2021learning} produce the query list using every word which occurs at least 100 times in English Wikipedia, plus bigrams from Wikipedia with high pointwise mutual information, the names of Wikipedia articles, and all WordNet synsets \cite{radford2021learning}.

\noindent\textbf{\textit{SLIP}} SLIP's training data is drawn from an English-language subset of the Yahoo Flickr Creative Commons (YFCC100M) dataset \cite{thomee2016yfcc100m}, and includes 15-million image-text pairs. YFCC100M includes more than 11 million images of people posted to the internet between 2002 and 2014 \cite{thomee2016yfcc100m}. Note that while SLIP outperforms the CLIP architecture when both models are trained on the 15-million image-text dataset, SLIP does not outperform the versions of CLIP trained on 400 million images \cite{mu2021slip}. This research evaluates a version of CLIP trained on 400 million images, but only has access to a version of SLIP trained on 15 million images; thus, results obtained from CLIP are expected to better reflect societal associations and biases.

\noindent\textbf{\textit{BLIP}} BLIP's training dataset consists of 14-million image-text pairs drawn from five datasets: COCO ("Common Objects in Context"), an in-context object detection dataset \cite{lin2014microsoft}; Visual Genome, a densely annotated language-image dataset to enable recognition of relationships between objects in images \cite{krishnavisualgenome}; Conceptual Captions, a language-image dataset designed to train automatic image captioning AI \cite{sharma2018conceptual}; Conceptual 12M, a language-image dataset which relaxes the data collection pipeline of Conceptual Captions to include more data for language-image training \cite{changpinyo2021conceptual}; and SBU captions, a collection of over a million images and captions collected and filtered from Flickr \cite{ordonez2011im2text}. Most BLIP models, including those examined in this research, also train on a subset of 115-million image-text pairs from the LAION-400M open source dataset, which is intended to imitate the WIT dataset  \cite{radford2021learning}, and which uses CLIP cosine similarity measurements to filter low-quality image-text pairs \cite{schuhmann2021laion}. \citet{birhane2021multimodal} found evidence of pornographic, misogynistic, and stereotypical images and text in LAION-400m \cite{schuhmann_2021}, 

\section{Approach and Experiments}

Three experiments evaluate the bias associating American identity with being White in multimodal language-and-image AI. First, an experiment tests the ability of multimodal spaces to encode biases and veridical demographic information using an image ranking algorithm. Second, adaptations of the WEAT and SC-WEAT are used to evaluate bias related to American identity in visual semantic embedding spaces. Third, bias related to American identity is assessed in three downstream language-and-image tasks: visual question answering, image captioning, and synthetic image generation.

\noindent\textbf{State-Level Correlations with Bias and Demographics} The correspondence of visual semantic representations with U.S. state-level demographics is assessed using an image ranking algorithm. Each of the 597 images of the CFD are embedded using the vision transformer of the multimodal model in question. The image embeddings are randomly balanced based on race or ethnicity to account for group size imbalances, such that each of the four races or ethnicities in the CFD has 108 embedded images, for 432 total. Then, an embedding reflecting the name of each state is obtained by providing the linguistic context "a photo of someone who lives in [STATE]", based on the prompting format suggested by \citet{radford2021learning} and adopted by \citet{mu2021slip}, 
to the text encoder of the multimodal model. The cosine similarity of the text embedding for each state is obtained with all 432 image embeddings, and the images are ranked by cosine similarity from greatest to least. The 108 highest cosine similarities are selected, and the number of images corresponding to each race or ethnicity in the CFD  are counted. Using the 108 highest cosine similarities allows a state to be entirely associated with a single race or ethnicity, if this is what the model returns. To adjust for the effects of randomly downsampling, this process is repeated $1,000$ times for each state, and the mean count of images returned for each race or ethnicity is obtained. Results are evaluated based on Pearson's $\rho$ of the percent of images returned and the percent population of each state for each race or ethnicity. Correspondence between census statistics and images returned for all four races or ethnicities are obtained by fitting a multivariate linear regression and reporting the $R^{2}$ coefficient.

After measuring correspondence with state demographics, the number of images of Black individuals for each state is evaluated against state-level measurements of implicit bias for White online IAT participants and for Black online IAT participants. The intention of this experiment is to quantify whether the prototypical racial association of a state predicts the biases of White and Black individuals living in that state, as in research finding that fear of losing prototypical status predicts bias in White Americans \cite{danbold2015no}.

\noindent\textbf{EAT and SC-EAT} The present research employs a language-image version of the WEAT, which will be referred to as the EAT (Embedding Association Test), because the test is not restricted to the modality of language, as in the WEAT. An EAT and three SC-EATs are used to evaluate biases in the visual semantic embedding spaces of CLIP, SLIP, and BLIP. As defined by \citet{caliskan2017semantics}, EAT and SC-EAT approaches require attribute groups $A$ and $B$ to be of equal size, and the EAT requires target groups $X$ and $Y$ to be of the same size. However, this research uses the images of the CFD in attribute and target groups, and the number of images included in the CFD for each race or ethnicity is unequal. When population sizes are unequal, Cohen's $d$ can be obtained using a pooled standard deviation, for which the formula is provided in the appendix. A $p$-value is obtained using a Welch's $t$-test, which does not assume equal variance or population size. \citet{cohen1992statistical} defined an effect size of $.2$ as small, $.5$ as medium, and $.8$ as large.

\noindent\textbf{\textit{We/They EAT}} In evaluating bias related to American identity, \citet{devos2005american} design IAT attribute groups to represent prototypical in-group status and out-group status. One such test involves a task which pairs faces of White individuals with a "We" attribute group and faces of Asian individuals with a "They" attribute group. The stimuli used by \citet{devos2005american} are as follows:

\noindent\textbf{We}: we, our, ourselves \noindent\textbf{They}: they, other, themselves

Unlike the IAT, the EAT requires sets of at least 8 word or image stimuli to ensure the statistical significance of the results. Thus, the We/They attribute word groups are expanded to include 8 words per group, which are close variations of the originals:

\noindent\textbf{We}: we, our, ourselves, ours, us, familiar, similar, here

\noindent\textbf{They}: they, their, themselves, theirs, them, other, others, there

Words referring to an individual, such as "I" and "me," are not included in the "We" group, as \citet{devos2005american} note that the attribute words are intended to represent a collective national identity and a collective other. For a language-image EAT, the We/They groups are attribute groups $A$ and $B$ in the EAT formula. The target group $X$ is a collection of image embeddings of White individuals in the CFD. The target group $Y$ is a collection of image embeddings of Asian individuals, Black individuals, or Latina/o individuals from the CFD. For this test, a positive effect size indicates greater similarity of in-group words with the White target group, and a negative effect size indicates greater similarity of in-group words with a Black, Asian, or Latino/a target group.

\noindent\textbf{\textit{American Trait SC-EAT}} \citet{devos2005american} survey study participants to identify the core traits connected with American identity. The results of the survey indicate that three traits are more central than any others surveyed: whether a person is patriotic; whether they are native to the U.S.; and whether they treat people of all races and backgrounds equally, commensurate with the explicit ideology of racial equality in the U.S. Association of these three traits with the races and ethnicities included in the CFD is tested using the SC-EAT, with the following three phrases used as the target embedding:

\noindent "a photo of someone who is patriotic"

\noindent "a photo of someone who is an immigrant to America"

\noindent "a photo of someone who treats people of all races and backgrounds equally"

For these tests, the attribute group $A$ is composed of image embeddings of White individuals and the attribute group $B$ is composed of image embeddings of Asian, Black, or Latina/o individuals. A positive effect size indicates association of the phrase with White individuals, while a negative effect size indicates association of the phrase with Asian, Black, or Latina/o individuals. These tests observe the "a photo of" prompting format described by \citet{radford2021learning} and used to obtain results which correspond well to census ground truth in the experiment described previously.

\noindent\textbf{Downstream Language-and-Image Tasks} This research evaluates the presence of biases related to American identity in three tasks for which language-and-image AI is widely used: visual question answering, image captioning, and synthetic image generation. 

\noindent\textbf{\textit{Visual Question Answering}} The BLIP visual question answering model is prompted with a text question: "Is this person an American?" Each of the images in the CFD is input to the model, which responds with an answer as to whether the person is an American. The number of times the model answers "yes" or "no" are quantified for each race or ethnicity in the CFD, and the percentage of the time the response was "yes" is calculated. While the BLIP visual question answering model is capable of generating answers other than "yes" and "no," and sometimes generates text such as "I don't know," the model responds either "yes" or "no" in all cases for this experiment. The BLIP visual question answering model is then provided with a second text question: "What state does this person live in?" Each of the images in the CFD is input to the model, and in all cases the model responds with a one-word answer. For each race or ethnicity, the answers are counted, and the percentage of the time an answer occurs for each race or ethnicity in the CFD is calculated.

\noindent\textbf{\textit{Image Captioning}} Each of the images in the CFD is input to the BLIP image captioning model, and captions are generated with the top-$p$ parameter set to .5, .6, .7, .8, and .9. Top-$p$ or "nucleus" sampling controls the randomness of the output of a text generator by selecting the next token from among the minimum number of tokens which make up $p * 100$\% of the probability mass for that word. Higher values of $p$ allow the model to select from a wider variety of sentence continuations, resulting in more varied output. At low values of $p$, such as below .5, the model is restricted to choosing from among a small subset of words, and in almost all cases generates a caption similar to "a photo of a man wearing a grey shirt" for the images of the CFD. Letting top-$p$ vary between .5 and .9 allows assessment of a wider variety of high-probability model outputs. At each level of the top-$p$ parameter, an automatically generated image is obtained for each of the CFD images. The percentage of the time a caption describes the race or ethnicity of the person in an image is calculated for each of the races or ethnicities in the CFD. Similar to the We/They EAT, this task serves as a way of measuring what races or ethnicities are prototypical, and not in need of description; and what races or ethnicities are other, and in need of description.

\begin{figure*}
    \includegraphics[width=5.5cm]{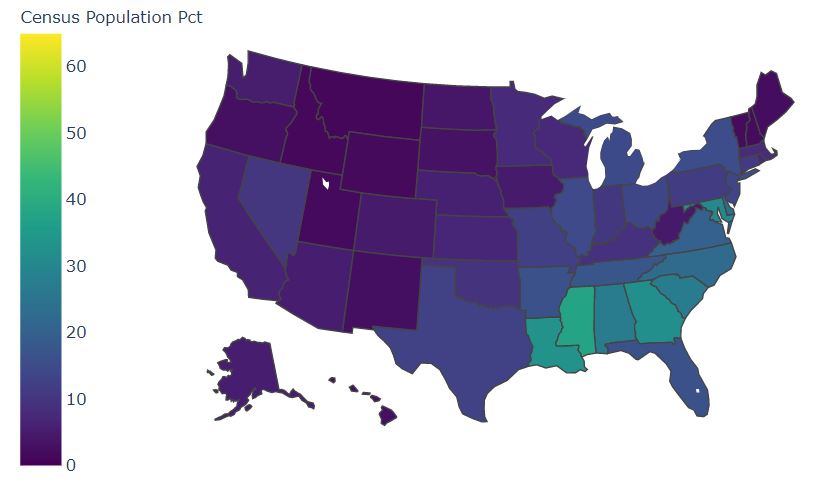}
    \includegraphics[width=5.5cm]{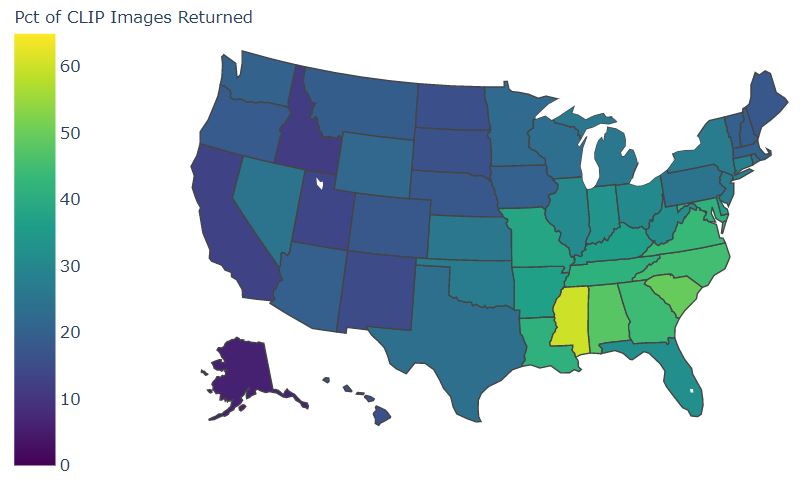}
    \includegraphics[width=5.5cm]{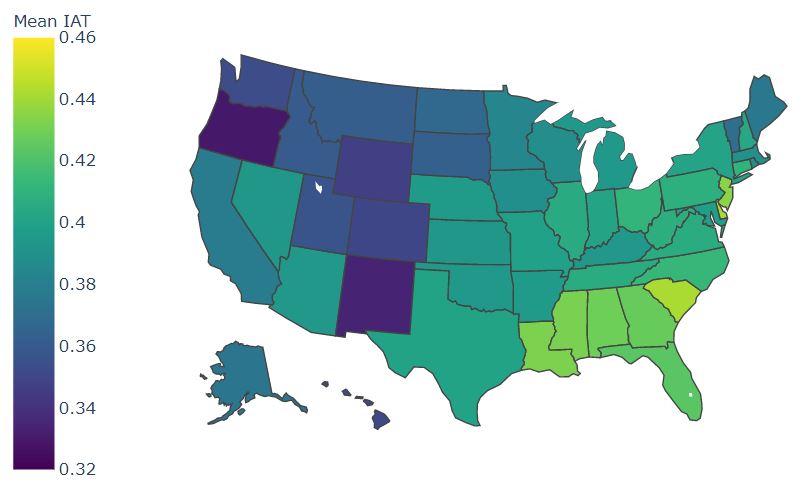}
    \caption{\small The number of images returned by CLIP (center image) when prompted with "a photo of someone who lives in [STATE]" correlates strongly with 2019 census figures (left image, same scale as center image, 0-65\%). However, CLIP overrepresents Black individuals relative to census statistics. Comparing to mean IAT scores for White individuals from each state (right image, min-max scaled) reveals that IAT scores correlate more strongly with national racial associations captured by AI ($\rho=.64$ in CLIP, $\rho=.69$ in BLIP) than with census statistics ($\rho=.60$.)}
    \label{census_ret_iat}
    \Description{Choropleths showing census population demographics, CLIP images retrieved, and mean IAT scores for White individuals.}
\end{figure*}

\noindent\textbf{\textit{Synthetic Image Generation}} The VQGAN-CLIP synthetic image generation model is provided with an initialization image, which is used as the starting point for the production of an output image. Every image in the CFD is used as an initialization image. The model is provided the text prompt "an American person" to guide the generation of a synthetic image. This experiment is performed using a checkpoint for WikiArt-16034, which is trained to generate synthetic artwork. While checkpoints exist for generating more photographic representations of faces, these checkpoints often generate unrealistic images missing facial features such as eyes or nose. This appears to be much less common with the WikiArt checkpoint.

For each of the generated images, a 50x50 pixel square of the image is cropped from the part of the face corresponding to the forehead just above the eyebrows and between the eyes. The mean brightness of the pixels in this square are measured as a proxy to understanding biases related to skin tone. This experiment assesses whether the bias associating American with White is strong enough that the model generates an image of a White individual even when provided with an initialization of a person who was recruited in America and who self-identifies as Asian, Black, or Latina/o. All synthetic images are trained for 200 iterations, the default setting of the model, and generated images are 592x592 pixels. The appendix includes additional notes concerning the design of this experiment.

\section{Results}
Results indicate that language-and-vision AI associates American with White, and that this bias manifests in downstream tasks.

\begin{table}[htbp]
\centering
{\small
\begin{tabular}
{|l|c|l|c|l|c|l|c|l|c|}
 \hline
 \multicolumn{10}{|c|}{Correlation of Retrieved Images with Census Population} \\
 \hline
 \multirow{2}{*}{Model} & \multicolumn{2}{c}{Asian} & \multicolumn{2}{|c|}{Black} & \multicolumn{2}{|c|}{Latina/o} & \multicolumn{2}{|c|}{White} & All\\
 \cline{2-10}
 &\multicolumn{1}{|c|}{$\rho$} & \multicolumn{1}{|c|}{$p < $} & \multicolumn{1}{|c|}{$\rho$} & \multicolumn{1}{|c|}{$p < $} & \multicolumn{1}{|c|}{$\rho$} & \multicolumn{1}{|c|}{$p < $} & \multicolumn{1}{|c|}{$\rho$} & \multicolumn{1}{|c|}{$p < $} & {$R^{2}$}\\
 \hline
   BLIP & \cellcolor{gray!72}$.72$  & $10^{-9}$  & \cellcolor{gray!70}$.70$  & $10^{-8}$  & \cellcolor{gray!58}$.58$  & $10^{-6}$  & \cellcolor{gray!71}$.71$  & $10^{-9}$ & \cellcolor{gray!54}$.54$ \\
   CLIP & \cellcolor{gray!75}$.75$  & $10^{-9}$  & \cellcolor{gray!90}$.90$ & $10^{-19}$  & \cellcolor{gray!81}$.81$  & $10^{-13}$  & \cellcolor{gray!84}$.84$  & $10^{-14}$ & \cellcolor{gray!74}$.74$ \\
   SLIP & \cellcolor{gray!43}$.43$  & $10^{-2}$ & \cellcolor{gray!42}$.42$  & $10^{-2}$  & \cellcolor{gray!57}$.57$  & $10^{-5}$  & \cellcolor{gray!53}$.53$  & $10^{-5}$ & \cellcolor{gray!34}$.34$ \\
 \hline
\end{tabular}
\caption{\small The embedding spaces of BLIP, CLIP, and SLIP associate U.S. states with images of individuals who belong to the races and ethnicities living in that state, according to census demographics. CLIP trains on more than 3 times as much data as the other models, and exhibits the strongest correlation between demographics and model associations, with $R^{2}=.74$. Correlations are Pearson's $\rho$.}
\label{tab:veridicality}}
\Description{Table displaying the correlation of demographic groups in images retrieved by BLIP, CLIP, and SLIP with census population statistics.}
\end{table}

\noindent\textbf{State-Level Correlations of Bias and Demographics} Results reported in Table \ref{tab:veridicality} indicate that strong correlations exist between the race or ethnicity of the highest ranked images for each state returned by BLIP, CLIP, and SLIP, and the demographics of that state as reported in 2019 U.S. Census data. Specifically, the $R^{2}$ coefficient obtained for CLIP is $.74$, for BLIP is $.54$, and for SLIP is $.33$. This corresponds to the amount of data on which the three models train, as CLIP trains on $400$ million image-text pairs, BLIP on $129$ million image-text pairs, and SLIP on $15$ million image-text pairs. Pearson's $\rho$ for individual races or ethnicities ranges between $.42$ and $.90$ with CLIP again achieving the highest correlations of any model, and SLIP the lowest. While these results indicate that real-world population correlates with the model's associations, a closer look at the data reveals that images of Asian and Black individuals are generally over-represented in image ranking results relative to census population, while White individuals are under-represented relative to census population. This suggests that biases associating White with American are not the straightforward result of White individuals being the most populous U.S. racial group, but also reflect attitudes related to identity, as this research shows.

For CLIP and BLIP, the association of a state with images of Black individuals correlates positively with the mean racial bias IAT score for White individuals residing in each state. While state demographics are also positively correlated, the results show that IAT scores for White individuals correlate more strongly with the percentage of images of Black individuals returned by language-and-image AI (Pearson's $\rho=.64$ in CLIP, and $\rho = .69$ in BLIP) than with the proportion of state population which is Black according to census figures ($\rho = .60$). This suggests a possible link between the degree to which the model associates a region with Black individuals, and the bias of White individuals who live in that region. Conversely, the percentage of images of Black individuals returned correlates negatively ($\rho = -.55$ in CLIP and $\rho = -.41$ in BLIP) with pro-White implicit bias scores for Black individuals. Figure \ref{census_ret_iat} visualizes the relationship between census statistics, CLIP biases, and IAT scores.

\noindent\textbf{EAT and SC-EAT Associations} Findings of bias using the EAT reflect those reported in human subjects by \citet{devos2005american}. White individuals are differentially associated with in-group words and with patriotism, while Asian, Black, and Latina/o individuals are differentially associated with egalitarianism and with not being native to America when compared with White individuals.

\noindent\textbf{\textit{We/They EAT}} Table 2 shows that in BLIP, CLIP, and SLIP, a We/They EAT associates White individuals with in-group or collective "We" words, and associates Asian individuals with out-group or "Other" words, with effect sizes ranging between $.46$ and $.64$, and $p$-values of $10^{-3}$ or smaller. In BLIP and SLIP, White is associated with the We group and Black with the They group, while in CLIP there is no statistically significant result for the White vs. Black EAT. In BLIP and CLIP, White is associated with the We group and Latina/o with the They group, while SLIP appears to reverse this.

\begin{table}[htbp]
\centering
{\small
\begin{tabular}
{|l|r|c|r|c|r|c|}
 \hline
 \multicolumn{7}{|c|}{We/They EAT} \\
 \hline
 \multirow{2}{*}{Model} & \multicolumn{2}{c}{White/Asian} & \multicolumn{2}{|c|}{White/Black} & \multicolumn{2}{|c|}{White/Latina/o}\\
 \cline{2-7}
 &\multicolumn{1}{|c|}{$d$} & \multicolumn{1}{|c|}{$p <$} & \multicolumn{1}{|c|}{$d$} & \multicolumn{1}{|c|}{$p <$} & \multicolumn{1}{|c|}{$d$} & \multicolumn{1}{|c|}{$p <$} \\
 \hline
   BLIP & \cellcolor{gray!51}$.51$  & $10^{-5}$  & \cellcolor{gray!34}$.34$  & $10^{-3}$  & \cellcolor{gray!65}$.65$  & $10^{-7}$ \\
   CLIP & \cellcolor{gray!46}$.46$  & $10^{-3}$  & $-.01$ & $n.s.$  & \cellcolor{gray!45}$.45$  & $10^{-3}$ \\
   SLIP & \cellcolor{gray!64}$.64$  & $10^{-8}$  & \cellcolor{gray!49}$.49$  & $10^{-6}$  & $-.49$  & $10^{-4}$ \\
 \hline
\end{tabular}
\caption{\small In the embedding spaces of BLIP, CLIP, and SLIP, images of White individuals are differentially associated with collective in-group words (we, our, ourselves), while Asian, Black, and Latina/o individuals are differentially associated with out-group words (they, other, themselves), suggesting that collective in-group identity is more readily associated with White individuals in language-and-image AI. Gray shading indicates strength of association with White.}
\label{tab:wetheyweat}}
\Description{Table displaying results for embedding association test of inclusive "We" words vs. othering "They" words.}
\end{table}

\begin{table}[htbp]
\centering
{\small
\begin{tabular}
{|l|r|c|r|c|r|c|}
 \hline
 \multicolumn{7}{|c|}{Patriotism SC-EAT} \\
 \hline
 \multirow{2}{*}{Model} & \multicolumn{2}{c}{White/Asian} & \multicolumn{2}{|c|}{White/Black} & \multicolumn{2}{|c|}{White/Latina/o}\\
 \cline{2-7}
 &\multicolumn{1}{|c|}{$d$} & \multicolumn{1}{|c|}{$p <$} & \multicolumn{1}{|c|}{$d$} & \multicolumn{1}{|c|}{$p <$} & \multicolumn{1}{|c|}{$d$} & \multicolumn{1}{|c|}{$p <$} \\
 \hline
   BLIP & \cellcolor{gray!56}$.56$  & $10^{-6}$  & \cellcolor{gray!32}$.32$  & $10^{-3}$  & \cellcolor{gray!29}$.29$  & $.01$ \\
   CLIP & \cellcolor{gray!28}$.28$  & $.05$  & \cellcolor{gray!52}$.52$ & $10^{-7}$  & \cellcolor{gray!62}$.62$  & $10^{-7}$ \\
   SLIP & \cellcolor{gray!35}$.35$  & $.01$  & \cellcolor{gray!100}$1.23$  & $10^{-28}$  & \cellcolor{gray!15}$.15$  & $n.s.$ \\
 \hline
\end{tabular}
\caption{\small The phrase "a photo of someone who is patriotic" is without exception differentially associated with White individuals in BLIP, CLIP, and SLIP, commensurate with the findings of \citet{devos2005american}, who found that White Americans are viewed as more patriotic. Gray shading indicates strength of association with White.}
\label{tab:patriotism}}
\Description{Table displaying results from a Patriotism embedding association test.}
\end{table}

\begin{table}[htbp]
\centering
{\small
\begin{tabular}
{|l|r|c|r|c|r|c|}
 \hline
 \multicolumn{7}{|c|}{Egalitarianism SC-EAT} \\
 \hline
 \multirow{2}{*}{Model} & \multicolumn{2}{c}{White/Asian} & \multicolumn{2}{|c|}{White/Black} & \multicolumn{2}{|c|}{White/Latina/o}\\
 \cline{2-7}
 &\multicolumn{1}{|c|}{$d$} & \multicolumn{1}{|c|}{$p <$} & \multicolumn{1}{|c|}{$d$} & \multicolumn{1}{|c|}{$p <$} & \multicolumn{1}{|c|}{$d$} & \multicolumn{1}{|c|}{$p <$} \\
 \hline
   BLIP & \cellcolor{gray!97}$-.97$  & $10^{-30}$  & \cellcolor{gray!100}$-3.10$  & $10^{-30}$  & \cellcolor{gray!100}$-1.09$  & $10^{-30}$ \\
   CLIP & \cellcolor{gray!100}$-1.31$  & $10^{-30}$  & \cellcolor{gray!100}$-1.96$ & $10^{-30}$  & \cellcolor{gray!37}$-.37$  & $.01$ \\
   SLIP & $.84$  & $10^{-12}$  & $1.05$  & $10^{-22}$  & $.05$  & $n.s.$ \\
 \hline
\end{tabular}
\caption{\small Egalitarianism is differentially associated with Asian, Black, or Latina/o individuals in BLIP and CLIP, indicating that, despite prototypical association with American identity, egalitarianism is not readily associated with White individuals, commensurate with the findings of \citet{devos2005american}. Shading reflects strength of association with Asian, Black, or Latina/o in the SC-EAT.}
\label{tab:egalitarianism}}
\Description{Table displaying results from an Egalitarianism embedding association test.}
\end{table}

\begin{table}[htbp]
\centering
{\small
\begin{tabular}
{|l|r|c|r|c|r|c|}
 \hline
 \multicolumn{7}{|c|}{Nativism SC-EAT} \\
 \hline
 \multirow{2}{*}{Model} & \multicolumn{2}{c}{White/Asian} & \multicolumn{2}{|c|}{White/Black} & \multicolumn{2}{|c|}{White/Latina/o}\\
 \cline{2-7}
 &\multicolumn{1}{|c|}{$d$} & \multicolumn{1}{|c|}{$p <$} & \multicolumn{1}{|c|}{$d$} & \multicolumn{1}{|c|}{$p <$} & \multicolumn{1}{|c|}{$d$} & \multicolumn{1}{|c|}{$p <$} \\
 \hline
   BLIP & \cellcolor{gray!100}$-1.03$  & $10^{-16}$  & \cellcolor{gray!100}$-1.50$  & $10^{-30}$  & \cellcolor{gray!100}$-1.38$  & $10^{-25}$ \\
   CLIP & \cellcolor{gray!100}$-1.19$  & $10^{-21}$  & \cellcolor{gray!22}$-.22$ & $.05$  & \cellcolor{gray!100}$-1.41$  & $10^{-23}$ \\
   SLIP & \cellcolor{gray!100}$-1.06$  & $10^{-20}$  & $.12$  & $n.s.$  & \cellcolor{gray!100}$-1.71$  & $10^{-30}$ \\
 \hline
\end{tabular}
\caption{\small The phrase "a photo of someone who is an immigrant to America" is differentially associated with Asian and Latina/o when compared with White in BLIP, CLIP, and SLIP. Effect sizes are small for a White vs. Black SC-EAT in CLIP and SLIP, reflecting findings that Black individuals are seen as native to the U.S., even if American identity is not readily conferred to them \cite{devos2005american,zou2017two}.  Shading reflects strength of association with Asian, Black, or Latina/o.}
\label{tab:nativism}}
\Description{Table displaying results for a Nativism embedding association test.}
\end{table}

\noindent\textbf{\textit{American Trait SC-EAT}} Table 3 indicates that, across all three visual semantic embedding spaces, White individuals are differentially associated with patriotism when compared with Asian, Black, or Latina/o individuals, with statistically significant effect sizes ($d$) ranging between $.28$ and $1.23$. Only one effect size is not statistically significant, for the White vs. Latina/o comparison in SLIP. Table 4 shows that, in BLIP and CLIP, Asian, Black, and Latina/o are strongly associated with egalitarianism when compared to White individuals, represented with the phrase "a photo of someone who treats people of all races and backgrounds equally." This finding does not hold for SLIP. Table 5 shows that, across all three models, Asian or Latina/o individuals are strongly associated with "a photo of someone who is an immigrant to America" when compared with White individuals. The effect holds for Black individuals vs. White individuals in BLIP, but not in CLIP or SLIP. The findings reflect those of \citet{devos2005american}, who found that White individuals are perceived to be more patriotic; that Asian individuals are perceived to have been born outside of the U.S.; and that White individuals are not as likely to be perceived as treating people of all races and backgrounds equally, despite the association of American identity with both egalitarianism and being White.

\noindent\textbf{Downstream Language-and-Image Tasks} Results indicate that biases associating White with American in language-and-image AI extend beyond visual semantic embedding spaces, and introduce bias into downstream tasks including visual question answering, image captioning, and synthetic image generation.

\begin{figure}[!htbp]
\begin{tikzpicture}
\begin{axis} [
    height=3cm,
    width=8cm,
    ybar = .05cm,
    bar width = 16pt,
    ymin = 0, 
    ymax = 100,
    ylabel=\% Answered Yes,
    ylabel near ticks,
    ylabel shift={-5pt},
    xtick = {1,2,3,4},
    xtick style={draw=none},
    ytick pos = left,
    xticklabels = {Asian, Latina/o, Black, White},
    x label style={at={(axis description cs:0.5,-0.1)},anchor=north},
    title=BLIP American Classifications by Race or Ethnicity,
    xlabel= {Self-Identified Race or Ethnicity},
    legend style={at={(0.37,0.73)},anchor=south west,nodes={scale=.82, transform shape}},
    enlarge x limits={abs=1cm}
]

\addplot [pattern=grid,pattern color = blue] coordinates {(1,2.7523) (2,61.1111) (3,68.5279) (4,96.7213)};

\end{axis}
\end{tikzpicture}
\vspace{-3mm}
\caption{\small The BLIP Visual Question Answering model identifies 96.7\% of White individuals as American, compared to only 2.8\% of Asian individuals. 68.5\% of Black individuals and 61.1\% of Latina/o individuals are identified as American.}
\label{fig:vqa_american}
\Description{Bar chart showing the BLIP VQA model's classifications of CFD images as American by race or ethnicity.}
\end{figure}
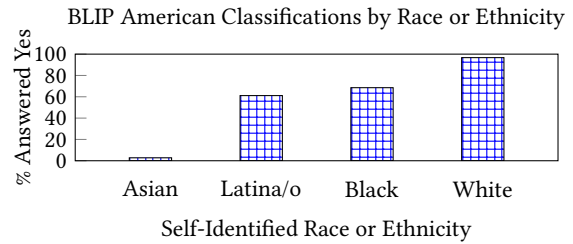

\noindent\textbf{\textit{Visual Question Answering}} As shown in Figure \ref{fig:vqa_american}, posing the question "Is this person an American?" to the BLIP visual question answering model yields unambiguous results: 96.7\% of White individuals are evaluated to be American, while only 2.8\% of Asian individuals are evaluated to be American. 68.5\% of Black individuals and 61.1\% of Latina/o individuals are evaluated to be American.

\begin{figure}
    \centering
    \includegraphics[width=.47\textwidth]{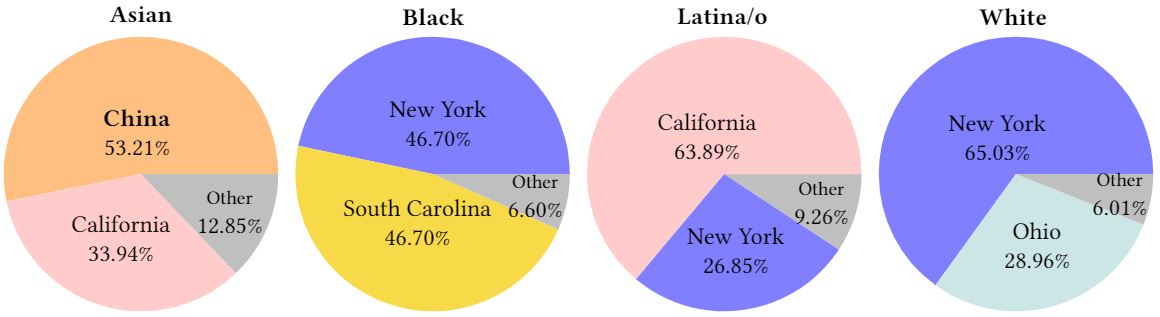}
\caption{\small Prompted with the question "What state does this person live in?", the BLIP Visual Question Answering head responds "China" for more than 53\% of images of Asian individuals, reflecting that the model fails to integrate Asian individuals into American identity. BLIP responds with a U.S. state for all White individuals.}
\label{fig:pie_charts}
\Description{Pie charts showing how the BLIP VQA head answers the question "What state does this person live in?" by race or ethnicity.}
\end{figure}

Posing the question, "What state does this person live in?" results in the model inferring that "state" refers to an American state for 100\% of images of White individuals, for which the model responds with New York 65.0\% of the time and Ohio 29.0\% of the time, as shown in Figure \ref{fig:pie_charts}. However, for images of Asian individuals, the model answers China 53.2\% of the time, reflecting that the bias toward Asian individuals is pronounced enough that the model does not associate the word "state" with an American state, which is the case for more than 97\% of the other images in the CFD. Similarly, the model answers that 3.1\% of Black individuals live in South Africa. New York seems to be a default state, and over-association with certain states occurs again in this experiment: BLIP answers South Carolina for 46.7\% of Black individuals, and California for 33.9\% of Asian and 63.9\% of Latina/o individuals.

\noindent\textbf{\textit{Image Captioning}} At every level of top-$p$ between $.5$ and $.9$ (the model has access to a greater number of sentence continuations as top-$p$ increases), the race of Asian individuals is the most commonly remarked upon by BLIP, with race noted between $12.84$\% and $36.70$\% of the time. The race of Black individuals is the second most remarked upon, between $2.03$\% of the time and $18.27$\% of the time. Race is never remarked upon for images of White individuals, and is only noted for Latina/o individuals when the model identifies an image of an individual who self-identified as Latina/o to be Asian or Black. While Black individuals are described as "African American" 92.9\% of the time when race is described, Asian individuals are always described as "Asian," and never as "Asian American."

While less common, the captions automatically generated by BLIP also reflect specific racial and gender associations and biases. In several troubling instances directly relevant to this research, Asian and Latina women are described as "oriental" in automatically generated captions, an offensive word directly highlighting the perceived foreignness of an individual \cite{said2014orientalism}. Latina/o individuals are sometimes described as "Asian," potentially indicating that the model forms a broad and not always distinguishable representation of racial outgroups. Moreover, generated captions sometimes describe Black individuals as having an afro or dreadlocks, when these individuals do not in fact have these hairstyles. Despite all of the photographs in the CFD being headshots, the generated captions sometimes remark on the chest size of women, and may describe whether they are wearing makeup, or if their skin is oily. Despite all subjects wearing an identical grey t-shirt visible only at the top of the shoulders, men are sometimes described as wearing a black tie, reflecting the association of men with professional environments. 

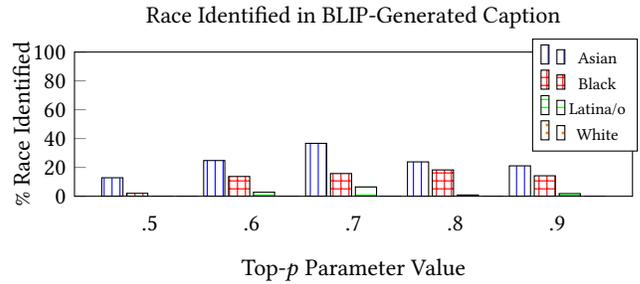
\begin{figure}[!htbp]
\begin{tikzpicture}
\begin{axis} [
    height=3.5cm,
    width=9cm,
    ybar = .05cm,
    bar width = 8pt,
    ymin = 0, 
    ymax = 100,
    ylabel=\% Race Identified,
    xtick = {1,2,3,4,5},
    xtick style={draw=none},
    ytick pos = left,
    ylabel near ticks,
    ylabel shift={-5pt},
    xticklabels = {.5, .6, .7, .8, .9},
    x label style={at={(axis description cs:0.5,-0.1)},anchor=north},
    title= Race Identified in BLIP-Generated Caption,
    xlabel= {Top-$p$ Parameter Value},
    legend style={at={(0.82,0.31)},anchor=south west,nodes={scale=.7, transform shape}},
    enlarge x limits={abs=1cm}
]

\addplot [pattern=vertical lines,pattern color = blue] coordinates {(1,12.84) (2,24.77) (3,36.70) (4,23.85) (5,21.1)};

\addplot [pattern=grid,pattern color = red] coordinates {(1,2.03) (2,13.71) (3,15.74) (4,18.27) (5,14.21)};

\addplot [pattern=horizontal lines,pattern color = green] coordinates {(1,0.00) (2,2.78) (3,6.48) (4,0.93) (5,1.85)};

\addplot [pattern=dots,pattern color = orange] coordinates {(1,0.00) (2,0.00) (3,0.00) (4,0.00) (5,0.00)};

\legend {Asian, Black, Latina/o, White};
\end{axis}
\end{tikzpicture}
\vspace{-3mm}
\caption{\small The race of Asian individuals is remarked upon by the BLIP image captioning model in 36.7\% of generated captions at Top-$p = .7$. The race of Black individuals is remarked upon by the BLIP image captioning model in 18.27\% of generated captions at Top-$p = .8$. The race of White individuals is never remarked upon, indicating default prototypical status in the model.}
\label{fig:image_captioning_american}
\Description{Bar chart showing the percentage of the time the BLIP image captioning head remarks on the race or ethnicity of an individual, with results broken down by race or ethnicity.}
\end{figure}

\begin{figure}[htbp]
\begin{tikzpicture}
\begin{axis} [
    height=4cm,
    width=9cm,
    line width = .5pt,
    ymin = 130, 
    ymax =220,
    xmin=-5.5,
    xmax=205.5,
    ylabel=Mean Pixel Brightness,
    ylabel shift=-5pt,
    ylabel near ticks,
    xtick = {0,10,20,30,40,50,60,70,80,90,100,110,120,130,140,150,160,170,180,190,200},
    xticklabel style={rotate=45,anchor=east},
    xtick pos=left,
    ytick pos = left,
    title=Change in Pixel Brightness of Forehead,
    xlabel= {Training Iteration},
    legend style={at={(.35,0.01)},anchor=south west,nodes={scale=.7, transform shape}}
]

\addplot [thick,solid,mark=*,color=red] coordinates {(0, 175.90829386060895) (10, 168.0608334702538) (20, 170.62576500362366) (30, 179.5250972135718) (40, 189.24381915282493) (50, 195.16036259102034) (60, 197.97903651256422) (70, 199.104747384818) (80, 199.27876381782815) (90, 198.5590968085418) (100, 197.5590917138246) (110, 196.4951409135083) (120, 194.75792750962816) (130, 193.53896138954448) (140, 192.04948941027652) (150, 190.76841053881873) (160, 189.14246893012685) (170, 188.43231343935912) (180, 187.60257300198992) (190, 185.94004041525912) (200, 185.4518746648631)};

\addplot [thick,densely dashed,mark=square*,color=violet] coordinates {(0, 134.83622447217775) (10, 133.4404425398047) (20, 141.60304469866264) (30, 154.53630502237152) (40, 166.4892920392025) (50, 174.99776390406984) (60, 179.88088930796602) (70, 181.63547901290426) (80, 181.8952988426283) (90, 181.39663183205528) (100, 180.9457107198002) (110, 179.8097662353038) (120, 178.95938945349198) (130, 177.96160088400842) (140, 176.67310035217753) (150, 175.83064241940008) (160, 174.4318445828977) (170, 173.61458519267762) (180, 172.62753260404534) (190, 171.6116562826713) (200, 171.12071359155905)};

\addplot [thick,dashed,mark=square*,color= orange] coordinates {(0, 177.64310523907218) (10, 188.4556216288481) (20, 197.44919886703494) (30, 205.94751797343008) (40, 211.6391508730137) (50, 213.27453514306728) (60, 213.8336025617571) (70, 212.96075363198878) (80, 212.1691892005166) (90, 210.91847605426463) (100, 208.90981614054942) (110, 206.8574375456767) (120, 205.78867033914068) (130, 204.38807205593056) (140, 202.4680243294853) (150, 200.97641312974935) (160, 199.47921725486498) (170, 197.37971599883696) (180, 196.14868935637404) (190, 194.76548686474976) (200, 193.382036834757)};

\addplot [thick,dotted,mark=triangle*,color=black] coordinates {(0, 185.71799622045762) (10, 197.2721187058519) (20, 203.38006097944927) (30, 209.81435179179286) (40, 213.77912284574805) (50, 214.91814354826352) (60, 214.50974487126086) (70, 213.9985567674031) (80, 211.90938528360434) (90, 210.2426160677373) (100, 208.5272396476913) (110, 206.90417053054165) (120, 205.02015427887244) (130, 202.9291033585607) (140, 200.92329853353002) (150, 199.3979173809562) (160, 198.2356561070702) (170, 196.74927957427195) (180, 194.4691706877813) (190, 192.91959228160763) (200, 191.4009593915365)};

\legend {{\small Asian,\small Black, \small Latina/o, \small White}};
\end{axis}
\end{tikzpicture}
\caption{ \small VQGAN-CLIP lightens skin tone, reflected in increased pixel brightness, when guided with the text "an American person." Provided with an initialization image of a Black individual, pixel brightness increases by 35\% on average. Regardless of initialization race or ethnicity, the model lightens skin tone.}
\Description{Figure showing the increase in the mean pixel brightness of the forehead (corresponding to the lightening of skin tone) by VQGAN-CLIP when guided with the text "an American person."}
\end{figure}
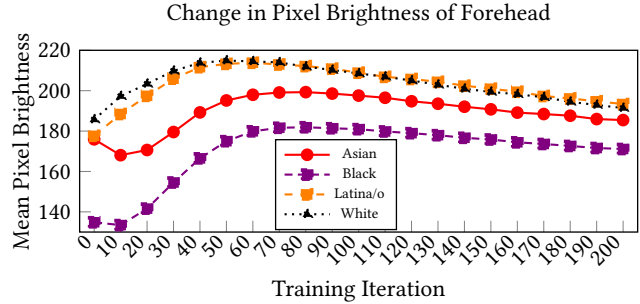

\noindent\textbf{\textit{Synthetic Image Generation}} In all cases, the CLIP-guided VQGAN increases the mean pixel brightness of the skin, indicating that skin tone has been lightened in response to the text "an American person." For Black individuals, there is a 35\% increase in mean pixel brightness over the first 80 training iterations, from 134.84 at the beginning of training to 181.89. However, the lightening of skin tone occurs even when the initialization image reflects an individual who self-identifies as White. Mean pixel brightness increases by 13\% for Asian individuals, by 20.4\% for Latina/o individuals, and by 15.7\% for White individuals at the highest brightness. Moreover, the data suggest that the generative model does not immediately locate the direction which lightens skin tone. In the first ten training iterations, the brightness of skin tone actually decreases for Asian and Black individuals. However, by the twentieth training step, the model finds that skin tone is the most impactful direction for matching the output image to the text input. Lightening of skin color continues until the eightieth step, when the model begins to add features such as American flags and bald eagles to the image. While the brightness of skin tone decreases after the one hundredth training step, qualitative inspection reveals that the model is not making skin tone darker; rather, the cropped region of the forehead tends to become obscured by an American flag, or by long blonde hair. Images later in the series appear to try to increase the realism of the depicted person by adding lines and shading to the forehead.

\section{Discussion}

Among the most cherished explicitly stated values of Americans are a commitment to the equality of all people without regard to race or ethnicity, and a belief that all people who make a home in America have an equal claim to American identity \cite{schuman1997racial,gross1999citizenship}. Yet expressed beliefs often do not reflect behavior, and experimental psychologists have found implicit attitudes excluding people who are not White from the category of American \cite{devos2005american}. This research demonstrates that these biases have been learned by AI, and that the implicit biases present in multimodal embedding spaces impact the biases which manifest in downstream language-and-image tasks.

The correlation between state-level IAT scores and the number of images of Black individuals returned by state in CLIP suggests that a connection exists between bias in human subjects and the threat to White prototypicality in a region, in keeping with the research of \citet{danbold2015no}, who find that fear of losing prototypical status predicts lower support for diversity among White Americans. The results suggest that, where a state is more associated on the national or international level with Black individuals than with White individuals, White racial biases are stronger. 

Tests assessing biases using the BLIP visual question answering head also suggest that American identity is not conferred upon any individual living in the U.S. While the model responds that only 61.1\% of Latina/o individuals are Americans, it also responds that 63.9\% of Latina/o individuals live in the state of California, and that 26.9\% of Latina/o individuals live in New York. Unlike for images for Asian individuals, for whom BLIP does not make the inference that an American state is being requested, BLIP is able to assign residence in an American state to Latina/o individuals - but not American identity. A similar result is observed for Black individuals, as the model responds that 68.5\% of Black individuals are American, but also responds that 46.7\% of Black individuals live in the state of South Carolina, and 46.7\% of Black individuals live in the state of New York. This incongruency suggests that to be American connotes something more in multimodal AI than living in the U.S. Moreover, that the model infers in most cases that the word "state" refers to an American territory, rather than to a nation-state, reflects an American-centric bias in the massive webscraped corpora used to train language-and-image AI.

Finally, a text-generation task renders the race of Asian, Black, and Latina/o individuals more visible than the race of White individuals, while an image-generation task renders visible only White individuals, even when provided with initialization images of Asian, Black, and Latina/o individuals. The BLIP image captioning model is more likely to draw attention to the race of Black individuals and especially Asian individuals, remarking upon the race of Asian individuals 36.7\% of the time when top-$p$ is set to $.7$. On the other hand, a CLIP-guided VQGAN lightens skin tone by at least 13\% based on pixel brightness for all races and ethnicities when provided with the text guidance "an American person," and by 35\% for initialization images of Black individuals. These results both derive from the default association of American identity with being White. Where only text referring to the category American is given, the most associated race is White. Where a model detects difference from the default, the deviation is described in racialized terms.

\begin{figure}
    \includegraphics[width=4cm]{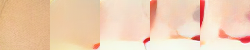}
    \includegraphics[width=4cm]{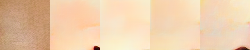}
    \includegraphics[width=4cm]{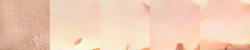}
    \includegraphics[width=4cm]{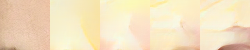}
    \caption{\small Ordered samples from the WikiArt 16384 checkpoint of VQGAN-CLIP for training iterations 0, 50, 100, 150, and 200 of images generated using the text "an American person." Initialization images are a self-identified Asian individual (top left), Black individual (top right), Latina/o individual (bottom left), and White individual (bottom right). Crops are of the forehead above the eyes. Skin tone is lightened, and American flags and forehead creases appear later.}
    \label{forehead_examples}
    \Description{Four series of sample images illustrating the lightening of skin tone of Asian, Black, Latina/o, and White individuals to maximize similarity with the target text "an American person."}
\end{figure}

This research suggests the potential impact of AI on humans, as racial prejudice related to exclusion from American identity has been shown to cause depression and low self-esteem \cite{armenta2013you,huynh2011perpetual}. Multimodal models similar to those examined in this research have been proposed as the future of ubiquitous internet applications such as Search \cite{nayak_2021}, and image generators capable of photorealistic output are expected to be widely usable by practitioners in the near future \cite{nichol2021glide}. Such systems are trained on internet-scale web scrapes, and are thus likely to encode societal biases consistent with those identified in this research. Unchecked, exposure to AI-generated content reflecting the prototypical association of American identity with being White may amplify a bias observed in humans.

\noindent\textbf{Limitations and Future Work} This research does not conduct a thorough evaluation of the training data for the models examined, and the training data for CLIP is not publicly available \cite{radford2021learning}. Previous work analyzing dataset composition has yielded insight into the causes of machine bias \cite{wolfe2021low,dodge2021documenting,birhane2021multimodal}, and future work might systematically explore the contents of those language-and-image AI training datasets which are publicly available. Moreover, this research examines three English-language models, because it is primarily concerned with whether an American bias is encoded into language-and-image AI. Should multilingual models comparable to CLIP be made available for research, they might be studied for the presence of similar ethnocentric biases. Additionally, language-and-image AI relies on language models to encode representations of text, which are inevitably sensitive to context. While this research defines prompts based on a principled method suggested by the designers of the models studied, it is unavoidable that changing the prompt will produce different embedding association results in some circumstances. Finally, future work might explore how other identity-related prototypicality associations manifest in AI.

\section{Conclusion}

In accordance with findings from experimental psychology \cite{devos2005american}, the present research reveals systematic biases in CLIP, SLIP, and BLIP associating American identity with being White. From biases associating White individuals with in-group words in visual semantic embedding spaces to the unambiguous exclusion of Asian, Latina/o, and Black individuals from American identity in visual question answering tasks, the results indicate that language-and-image AI learns that the prototypical American is White. 

\section*{Acknowledgements} This material is based on research partially supported by the U.S. National Institute of Standards and Technology (NIST) Grant \\60NANB20D212T. Any opinions, findings, and conclusions or recommendations expressed in this material are those of the authors and do not necessarily reflect those of NIST.

\bibliographystyle{ACM-Reference-Format}
\bibliography{sample-base}

\appendix

\section{Appendix}

\subsection{Foundational Research in Language-and-Image AI}

The ideas central to the design of modern language-and-image AI, including the use of natural language supervision to learn visual semantic features, were first introduced by \citet{socher2013zero}, and advanced by \citet{frome2013devise} in the design of the DeViSE deep learning visual semantic embedding model. While foundational for visual semantics, these models did not approach the performance of CLIP, which improved the zero-shot state-of-the-art on the ImageNet benchmark \cite{deng2009imagenet} from 11.5\% \cite{li2017learning} to 76.2\% \cite{radford2021learning}. The contrastive learning objective was introduced by \citet{tian2019contrastive}, and CLIP builds most directly on the ConVIRT medical image classifier of \citet{zhang2020contrastive}, and was designed concurrently with the zero-shot ALIGN language-image model of \citet{jia2021scaling}. Most recently, \citet{tiwary_2021} introduce Turing-NLG, a multilingual visual semantic model that outperforms CLIP on zero-shot image classification.\footnote{The models of \citet{jia2021scaling} and \citet{tiwary_2021} are not publicly available to researchers.} Recent research suggests that visual semantic pretraining may have benefits for language representations, independent of image representations, as the word and sentence embeddings formed by CLIP have been shown to be highly semantic \cite{wolfe2022contrastive}, setting or matching state of the art on the RG65 \cite{rubenstein1965contextual} and ValNorm \cite{toney2020valnorm} intrinsic evaluations.

\subsection{WEAT Formulae}

As described by \citet{caliskan2017semantics}, the formula for the WEAT is given by:

\begin{equation}
\frac{\textrm{mean}_{x\in X}s(x,A,B) - \textrm{mean}_{y\in Y}s(y,A,B)}{\textrm{std\_dev}_{w \in X \cup Y}s(w,A,B)}
\end{equation}

where the association for a word w is:
 
\begin{equation}
{\textrm{mean}_{a\in A}\textrm{cos}(\vec{w},\vec{a}) - \textrm{mean}_{b\in B}\textrm{cos}(\vec{w},\vec{b})}
\end{equation}

The SC-WEAT measures the differential angular similarity between two groups of attribute words with a single target word. The formula for the SC-WEAT is given by:

\begin{equation}
\frac{\textrm{mean}_{a\in A}\textrm{cos}(\vec{w},\vec{a}) - \textrm{mean}_{b\in B}\textrm{cos}(\vec{w},\vec{b})}{\textrm{std\_dev}_{x \in A \cup B}\textrm{cos}(\vec{w},\vec{x})}
\end{equation}

When population sizes are unequal, Cohen's $d$ can be obtained using a pooled standard deviation, given by:

\begin{equation}
    \sigma_{p} = \sqrt{\frac{(N_{1}-1) \textrm{std\_dev}_{1}^{2} + (N_{2}-1) \textrm{std\_dev}_{2}^{2}}{N_{1} + N_{2} - 2}}
\end{equation}

\noindent where $N_{1}$ refers to the size of the first group, and $N_{2}$ to the size of the second group. 

Results from pooled standard deviation are broadly consistent with those obtained by randomly downsampling to equalize the size of the groups. In some cases, where population means are highly distinct and standard deviations are small, the pooled standard deviation leads to larger effect sizes than those obtained by downsampling.

\begin{figure}[htbp!]
    \centering
    \includegraphics[width=5.5cm]{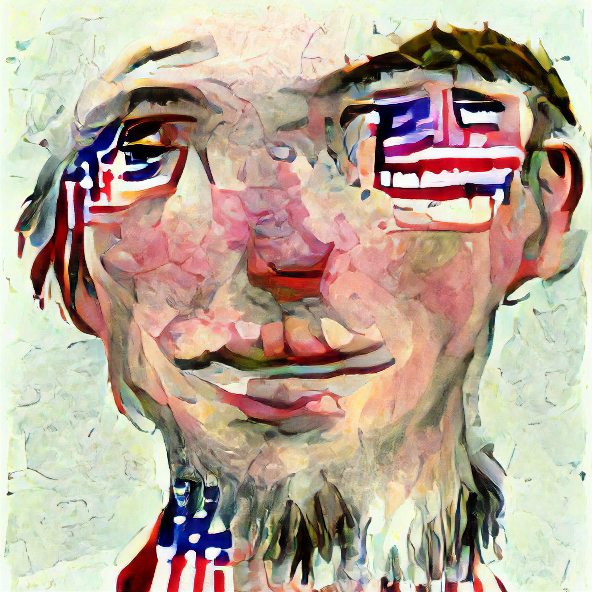}
    \includegraphics[width=5.5cm]{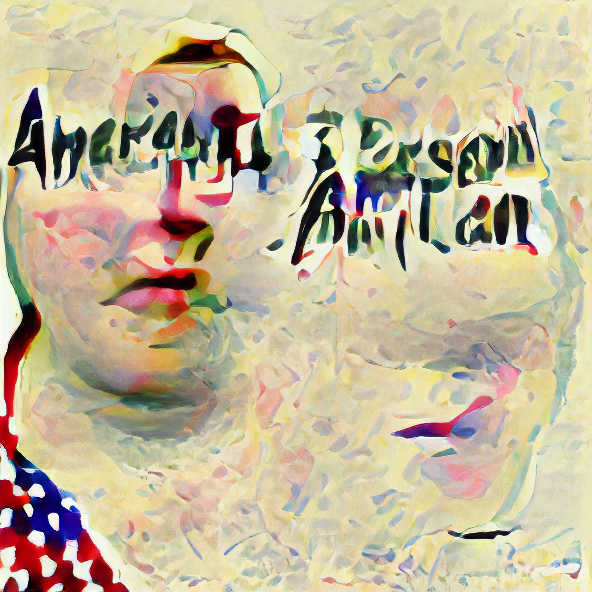}
    \includegraphics[width=5.5cm]{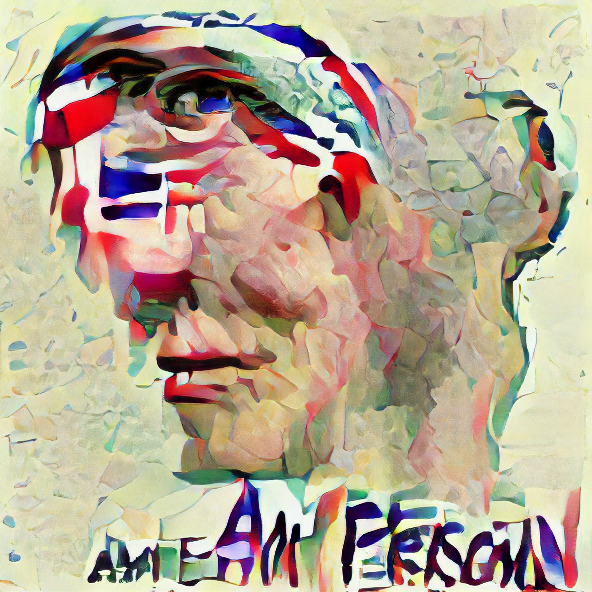}
    \caption{Images generated from the "an American person" text prompt with no initialization image, using only text guidance. Commensurate with the research of \citet{goh2021multimodal}, the model often generates a visual depiction of the input text.}
    \label{fig:text_only_guidance}
    \Description{Example of three images generated to match the text "an American person" with no initialization image provided.}
\end{figure}

\subsection{Experimental Design for Synthetic Image Generation}

The initial design of the synthetic image generation experiment included an additional experiment which generated images solely from the text "an American person," with no initialization image. However, provided only with text, the model tends to generate images of the word American, of flags, of bald eagles, and of other American symbols, and to create images of humans in inconsistent parts of the frame, making skin tone hard to measure. Qualitatively, we observe that where the generation of a human or humanlike image did occur, the individual was White, and in most cases blonde. However, given the inconsistent results of early tests of this experiment, this research can make no generalizable quantitative claim regarding the use solely of text with no initialization image. Figure \ref{fig:text_only_guidance} provides three examples of synthetic images trained for 200 iterations and generated solely using the text input "an American person." The authors of this research used a publicly available, pretrained implementation of VQGAN-CLIP currently available at \url{https://colab.research.google.com/drive/1ZAus_gn2RhTZWzOWUpPERNC0Q8OhZRTZ}.

\end{document}